\documentclass[superscriptaddress,twocolumn,showpacs,preprintnumbers,amsmath,amssymb,reprint,showkeys]{revtex4-1}

\usepackage{stmaryrd}
\usepackage{txfonts}
\usepackage{mathrsfs}
\usepackage{graphicx}
\usepackage{dcolumn}
\usepackage{bm}
\usepackage{epsfig}
\usepackage{color}

\usepackage[colorlinks=true, linkcolor=blue, citecolor=blue, urlcolor=blue]{hyperref}
\begin{document}

\title{Switching dynamics of the spin density wave in superconducting $\mathbf{CeCoIn_5}$}

\author{Duk Y. Kim}
\affiliation{MPA-CMMS, Los Alamos National Laboratory, Los Alamos, New Mexico 87545, USA}
\author{Shi-Zeng Lin}
\email{szl@lanl.gov}
\affiliation{Theoretical Division, Los Alamos National Laboratory, Los Alamos, New Mexico 87545, USA}
\author{Eric D. Bauer}
\affiliation{MPA-CMMS, Los Alamos National Laboratory, Los Alamos, New Mexico 87545, USA}
\author{Filip Ronning}
\affiliation{MPA-CMMS, Los Alamos National Laboratory, Los Alamos, New Mexico 87545, USA}
\author{J. D. Thompson}
\affiliation{MPA-CMMS, Los Alamos National Laboratory, Los Alamos, New Mexico 87545, USA}
\author{Roman Movshovich}
\email{roman@lanl.gov}
\affiliation{MPA-CMMS, Los Alamos National Laboratory, Los Alamos, New Mexico 87545, USA}

\begin{abstract}
The ordering wave vector $\mathbf{Q}$ of a spin density wave (SDW), stabilized within the superconducting state of $\mathrm{CeCoIn_5}$ in a high magnetic field, has been shown to be hypersensitive to the direction of the field. $\mathbf{Q}$ can be switched from a nodal direction of the $d$-wave superconducting order parameter to a perpendicular node by rotating the in-plane magnetic field through the antinodal direction within a fraction of a degree. Here, we address the dynamics of the switching of $\mathbf{Q}$. We use a free energy functional based on the magnetization density, which describes the condensation of magnetic fluctuations of nodal quasiparticles, and show that the switching process includes closing of the SDW gap at one $\mathbf{Q}$ and then reopening the SDW gap at another $\mathbf{Q}$ perpendicular to the first one. The magnetic field couples to $\mathbf{Q}$ through the spin-orbit interaction. Our calculations show that the width of the hysteretic region of switching depends linearly on the deviation of magnetic field from the critical field associated with the SDW transition, consistent with our thermal conductivity measurements. The agreement between theory and experiment supports our scenario of the hypersensitivity of the $Q$ phase on the direction of magnetic field, as well as the magnon condensation as the origin of the SDW phase in $\mathrm{CeCoIn_5}$.
\end{abstract}

\date{\today}
\maketitle

\noindent
\textbf{Introduction} --
Magnetism and superconductivity represent two central themes of modern condensed-matter-physics research. In itinerant systems, both magnetism and superconductivity compete for the electronic density of state at the Fermi surface. This implies a route to induce superconductivity by suppressing magnetism by pressure, chemical doping, etc., and vice versa; while in systems with localized magnetic moments, the magnetic scattering of electrons is detrimental to the Cooper-pair formation. Therefore, it is widely believed that magnetism and superconductivity are antagonistic with each other. In the past decades, however, it has been found that superconductivity and magnetism can coexist microscopically in some compounds. The coexistence and interplay of superconductivity and magnetism poses a grand challenge to our understanding of these two phenomena and continues to be an active area of research.

$\mathrm{CeCoIn_5}$ is a prototypical heavy-fermion superconductor with a tetragonal crystal structure \cite{petrovic_heavy-fermion_2001,thompson_progress_2011}. It has a superconducting transition temperature $T_c = 2.3$ K at ambient pressure into a state with a $d_{x^2-y^2}$ pairing symmetry. At low temperature, the superconducting upper critical magnetic field $H_{c2}$ is mainly determined by strong Pauli pair breaking. Because of these unique properties, $\mathrm{CeCoIn_5}$ has been considered as a candidate \cite{PhysRevLett.89.137002,PhysRevLett.91.187004,radovan_magnetic_2003} for the long sought Fulde-Ferrell-Larkin-Ovchinnikov (FFLO) state \cite{PhysRev.135.A550,larkin_inhomogeneous_1965}. Extensive experimental measurements have revealed a new phase inside the superconducting state in the presence of a strong magnetic field. Later, the new phase was shown to be a spin-density-wave (SDW) order with two possible propagating wave vectors $\mathbf{Q}_{1,2}=(0.44,\ \pm 0.44,\ 0.5)$ by neutron scattering \cite{kenzelmann_coupled_2008,PhysRevLett.104.127001} and NMR measurements \cite{PhysRevLett.98.036402}. The direction of $\mathbf{Q}_{1,2}$ coincides with the nodal directions of the $d_{x^2-y^2}$ superconducting state. The magnitude of the moment is $0.15\ \mu_B$, with $\mu_B$ the Bohr magneton, and the moment is aligned along the crystallographic $c$ axis. One remarkable feature about this SDW phase is that the SDW phase is induced by an in-plane magnetic field of order of $10$ T and exists \emph{only inside} the superconducting phase, disappearing together with the superconductivity at $H_{c2}$.

Several theoretical proposals for the origin of the SDW phase have been put forward. It was argued that the vortex lattice enhances the density of state in the nodal direction of the $d$-wave pairing symmetry and triggers the formation of the SDW phase \cite{PhysRevB.83.140503}. It was also suggested that the coupling between the SDW and $d$-wave superconductivity leads to a pair density wave and/or FFLO that is responsible for the stabilization of the SDW phase \cite{PhysRevLett.102.207004,yanase_antiferromagnetic_2009,miyake_theory_2008,PhysRevLett.104.216403}. Pauli pair breaking can also stabilize the SDW in $\mathrm{CeCoIn_5}$ \cite{PhysRevB.82.060510,PhysRevB.83.224518}. It was suggested that the Zeeman splitting by a magnetic field creates Fermi pockets around the nodal directions, which promotes the nesting between quasiparticles and stabilizes the SDW order \cite{PhysRevLett.107.096401,PhysRevB.86.174517,PhysRevB.92.224510}.
Another related proposal is that the $d$-wave pairing symmetry enhances the magnetic susceptibility of the quasiparticle in the nodal direction when a magnetic field is applied \cite{PhysRevB.84.052508}. When the magnetic field reaches a threshold value, such that the magnetic susceptibility at wavevector $\mathbf{q}$ obtained in the random phase approximation $\chi(\mathbf{q})=\chi_0(\mathbf{q})/[1-U(\mathbf{q}) \chi_0(\mathbf{q})]$ diverges \cite{PhysRevB.84.052508}, the SDW phase is stabilized. Here, $U(\mathbf{q})$ is the interaction and $\chi_0(\mathbf{q})$ is the bare susceptibility. In this picture, there exist abundant magnetic fluctuations (magnons) centered at the wave vector $\mathbf{Q}_{1,2}$ in the superconducting phase. These magnons become soft upon increasing magnetic field and condense at the critical field when $\mathrm{Re}[\chi_0(\mathbf{q})U(\mathbf{q})]=1$. This magnon condensation picture is supported by recent neutron-scattering data \cite{PhysRevLett.100.087001,PhysRevLett.115.037001,PhysRevLett.109.167207}. These observations indicate that CeCoIn$_5$ is close to the SDW instability.

\begin{figure}[t]
\psfig{figure=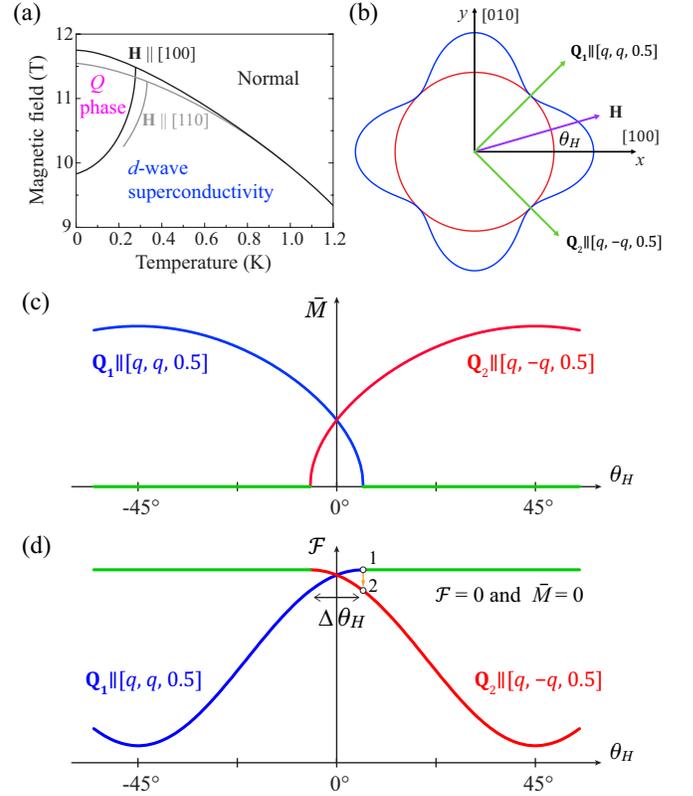,width=\columnwidth}
\caption{(color online) (a) The phase diagram of $\mathrm{CeCoIn_5}$ in the in-plane magnetic field \cite{PhysRevLett.91.187004}. (b) Schematic view of the magnetic-field direction and the two SDW ordering wave vectors. The system favors the SDW state with $\mathbf{Q}$ being more perpendicular to $\mathbf{H}$, while $\mathbf{Q}$ points along the nodes of the $d$-wave order parameter represented by the blue curve. The red circle denotes the normal Fermi surface. (c) Magnetic moment $\bar{M}_{1,2}$ and (d) free energy density $\mathcal{F}_{1,2}$ as a function of field angle $\theta_H$. The green line represents a nonmagnetic state, where the SDW gap vanishes. Here $q=Q_{1,2}/\sqrt{2}$.
} \label{f1}
\end{figure}

When the SDW forms via magnon condensation, there are two degenerate propagating vectors $\mathbf{Q}_{i}$ guaranteed by the $d$-wave pairing symmetry. The in-plane magnetic field breaks the two-fold degeneracy through spin-orbit coupling and selects one $\mathbf{Q}$. This is indeed observed by neutron scattering \cite{gerber_switching_2014}. $\mathbf{Q}$ changes sharply when one rotates the in-plane magnetic field. For instance, when the magnetic field is rotated from $[1\bar{1}0]$ to $[110]$, $\mathbf{Q}$ changes sharply from $\mathbf{Q}_1||[110]$ to $\mathbf{Q}_2||[1\bar{1}0]$ when the magnetic field rotates through the $[100]$ direction, as sketched in Fig. \ref{f1}(b). The hysteretic window is only about $0.3^\circ$ at $\mu_0 H\approx 11$ T \cite{gerber_switching_2014}. The thermal conductivity within the $Q$ phase in a rotating magnetic field reflected sharp switching of $\mathbf{Q}$, with a similar hysteresis \cite{Kim2016}. Phenomenologically, these observations suggest a coupling of the form $(\mathbf{Q}\times \mathbf{H})^2$ in the free energy functional \cite{mineev_antiferromagnetic_2015}. These two experiments also suggested the existence of a superconducting pair density wave within the $Q$ phase, in order to account for all the experimental observations \cite{gerber_switching_2014,Kim2016}. 

The transition between the two SDW states with different $\mathbf{Q}$ is of the first order according to Landau's argument. The first-order nature of the switching of different $\mathbf{Q}$ of the SDW manifests itself in a hysteresis, which has been confirmed experimentally \cite{gerber_switching_2014,Kim2016}. In a conventional first-order phase transition, the order emerges through nucleation of ordered domains with a finite correlation length. In contrast, neutron-scattering measurements  \cite{gerber_switching_2014} indicate that the SDW state switches as a whole, without the appearance of domains. In this Commnication, we argue that the switching of SDW occurs by closing the SDW gap at one $\mathbf{Q}$, when the barrier between the $\mathbf{Q}_{1}$ and $\mathbf{Q}_{2}$ states becomes zero, and then reopening the SDW gap at another (perpendicular) $\mathbf{Q}$. Such a process results in a hysteresis in switching, which increases linearly with the magnetic field according to our phenomenological model. Measurements of the width of the hysteresis region as a function of magnetic field are in agreement with the theoretical results. Our results corroborate the picture that the SDW phase is a consequence of the magnon condensation.\\ 

\noindent
\textbf{Phenomenological model} --
Near the phase transition, the SDW phase admits a Landau description based on a local order parameter $M_z(\mathbf{r})$. Because the moments align antiferromagnetically between layers of Ce atoms, it is sufficient to consider the magnetization inside one layer. The total free energy density near the low-field phase boundary of the SDW phase [see Fig. \ref{f1}(a)] can be written as 
\begin{align}\label{eq1}
\begin{split}
{\cal F} = - \frac{\alpha }{2}M_z^2 + \frac{\beta }{4}M_z^4 - \gamma {\left( {{\nabla _{2d}}{M_z}} \right)^2} \\
+ \eta \left[ {{{\left( {\partial _x^2{M_z}} \right)}^2} + {{\left( {\partial _y^2{M_z}} \right)}^2}} \right] 
- \lambda {\left[ {(\mathbf{H} \times {{\nabla _{2d}}){M_z}} } \right]^2},
\end{split}
\end{align}
where $\nabla_{2d}\equiv(\partial_x,\ \partial_y)$. The coupling between superconductivity and magnetism is taken into account through the coefficients, which depend on the superconducting order parameter. The term $[(\mathbf{H}\cdot\nabla_{2d})M_z]^2$ can be absorbed into the $\gamma$ and $\lambda$ terms, and, therefore, is not included in Eq. \eqref{eq1}. Zeeman coupling ($\mathbf{M}\cdot\mathbf{H}$) for the ordered moment is absent for an in-plane magnetic field because the ordered magnetic moments are along the $c$ axis in $\mathrm{CeCoIn_5}$. The $\eta$ term accounts for the anisotropy in ordering wave vector $\mathbf{Q}$. Experimentally $\mathbf{Q}||[110]$ or $\mathbf{Q}||[1\bar{1}0]$ indicating that $\eta>0$. The switching of the SDW domain suggests a coupling between $\mathbf{Q}$ and the magnetic field, which can originate from the spin-orbit interaction. This coupling is described by the $\lambda$ term, which lifts the degeneracy between the SDW solutions with $\mathbf{Q}_1||[110]$ and $\mathbf{Q}_2||[1\bar{1}0]$. The $\mathbf{Q}$ of the SDW prefers to align perpendicular to $\mathbf{H}$ when $\lambda>0$. This term was derived from a microscopic model of a two-band paramagnetic metal \cite{mineev_antiferromagnetic_2015}. We assume a weak coupling between field and $\mathbf{Q}$, $0<\lambda H^2\ll \gamma$. 

(a) We start with a single-${Q}$ SDW solution.
The magnetic moment arrangement in the SDW phase can be described by $M_z=\bar{M}\sin(\mathbf{Q}\cdot\mathbf{r})$. The corresponding free energy density is
\begin{align}\label{eq2}
\begin{split}
{{\cal F}} = - \left[ \frac{\alpha }{4} + \left[ {\frac{\gamma }{2} + \frac{\lambda }{2}{H^2}{{\sin }^2}\left( {{\theta _H} - \phi } \right)} \right]{Q^2} \right. \\ \left.
- \frac{{\eta {Q^4}}}{2}\left( {1 - \frac{{{{\sin }^2}\left( {2\phi } \right)}}{2}} \right) \right]\bar{M}^2 + \frac{{3\beta }}{{32}}\bar{M}^4,
\end{split}
\end{align}
where $\phi$ ($\theta_H$) is the angle between $\mathbf{Q}$ ($\mathbf{H}$) and the $x$ axis. The optimal $\phi$ to  linear order in $\lambda$ is
\begin{align}\label{eq3}
\phi_{1,2}=\pm\frac{\pi}{4}+\frac{{{H^2}\lambda \cos \left( {2{\theta _H}} \right)}}{{4\gamma }}.
\end{align}
corresponding to $\mathbf{Q}_1$ and $\mathbf{Q}_2$ in Fig. \ref{f1}(b) with a small correction due to the $\lambda$ term. The optimal $Q$ is
\begin{align}\label{eq4}
Q_{1,2}^2 = \frac{{ {2\gamma + {H^2}\lambda [1 \mp \sin \left( {2{\theta _H}} \right)]} }}{{{2\eta } }}.
\end{align}
Both $\phi_{1,2}$ and $Q_{1,2}$ receive a small correction of the order of $\lambda H^2/\gamma\ll 1$ from the spin-orbit coupling. 
The magnitude of the modulation $\bar{M}_1$ is
\begin{align}\label{eq5}
\bar{M}_{1}=2\sqrt{\frac{ { {\alpha \eta + {\gamma ^2}}  + {H^2}\gamma \lambda \left[ {1 - \sin \left( {2{\theta _H}} \right)} \right]} }{ {3\beta \eta } }}.
\end{align}
when $\sin(2\theta_H) \le ({{{\gamma ^2} + \alpha \eta + {H^2}\gamma \lambda }})/{{{H^2}\gamma \lambda }}$ and $\bar{M}_{1}=0$ otherwise. For $\bar{M}_2$, we have
\begin{align}\label{eq6}
\bar{M}_{2}=2\sqrt{\frac{ { {\alpha \eta + {\gamma ^2}}  + {H^2}\gamma \lambda \left[ {1 + \sin \left( {2{\theta _H}} \right)} \right]} }{ {3\beta \eta } }}.
\end{align}
when $\sin(2\theta_H) \ge -( {{{\gamma ^2} + \alpha \eta + {H^2}\gamma \lambda }})/{{{H^2}\gamma \lambda }}$ and $\bar{M}_{2}=0$ otherwise. 
The corresponding free energy for the SDW with $\mathbf{Q}_1$ and $\mathbf{Q}_2$ is
\begin{align}\label{eq7}
\mathcal{F}_{1,2}= -\frac {\left[ { {\alpha \eta + {\gamma ^2}}  + {H^2}\gamma \lambda \left[ {1 \mp \sin \left( {2{\theta _H}} \right)} \right]} \right]^2} { {6\beta {\eta ^2}}},
\end{align}
when $\bar{M}_{1,2}>0$, and $\mathcal{F}_{1,2}=0$ when $\bar{M}_{1,2}=0$.

From Eqs. \eqref{eq5} and \eqref{eq6}, it is clear that the critical field $H_{\mathrm{SDW}}$ of the SDW transition depends on the field angle $\theta_H$ because of the $\lambda$ term. In addition, the superconducting properties change with the field angle, as manifested by the change in $H_{c2}$ for fields along $[100]$ and $[110]$, see Fig. \ref{f1}(a). The effect of superconductivity is accounted for by the coefficients in $\mathcal{F}$ in Eq. \eqref{eq1}. Therefore, there is an intrinsic dependence of $H_{\mathrm{SDW}}$ on the field angle through $\alpha$, $\beta$, $\eta$ and $\gamma$. This makes the experimental determination of the dependence of $H_{\mathrm{SDW}}$ on $\theta_{H}$ due to the spin-orbit coupling difficult. The field dependence of $\bar{M}_{1,2}$ is $\bar{M}_{1,2}^2\propto {H-H_{\mathrm{SDW}}(\theta_H)}$. The linear dependence of $\bar{M}_{1,2}^2$ on $H$ is consistent with neutron-scattering data \cite{gerber_switching_2014}. The effective dimension of the quantum phase transition at $H_0'$ is $D'=D+z$, which is greater than the upper critical dimension. This renders the transition mean-field type. Here, $z$ is the dynamic critical exponent, and $D$ is the physical dimension. 

The switching behavior is determined by $G\equiv({{{\gamma ^2}+\alpha \eta + {H^2}\gamma \lambda }})/{{{H^2}\gamma \lambda }}$, which has four distinct cases described below. $\mathrm{CeCoIn_5}$ corresponds to the first case with $0<G\ll 1$. $G$ may be tuned by magnetic field, pressure and chemical doping etc.

1. For $0< G<1$, relevant for $\mathrm{CeCoIn_5}$, an illustration of $\mathcal{F}_i(\theta_H)$ for $\mathbf{Q}_1$ and $\mathbf{Q}_2$ according to Eq. \eqref{eq7} is shown in Fig. \ref{f1}(d). The SDW with $\mathbf{Q}_1$ is favored when $-90^\circ \le \theta_H\le0^\circ$ and the SDW with $\mathbf{Q}_2$ is more stable when $0^\circ \le \theta_H\le 90^\circ$. When the magnetic field rotates in the $ab$ plane and an increasing $\theta_H$ passes through $\theta_H = 0$, it is not possible for the SDW to change continuously from $\mathbf{Q}_1$ to $\mathbf{Q}_2$ because of the energy barrier presented by the $d$-wave order parameter. We argue that the switching of SDW $\mathbf{Q}$ is accomplished by complete suppression of the SDW gap at $\mathbf{Q}_1$ [point 1 in Fig. \ref{f1}(d)] and then reopening the gap at $\mathbf{Q}_2$  [point 2 in Fig. \ref{f1}(d)], see also Fig. \ref{f1} (c) for $\bar{M}$. This dynamic process is hysteretic. The field angle in the vicinity of $[100]$ at which the gap is completely suppressed is
\begin{align}\label{eq8}
\sin \left( {2{\theta _H}} \right) = \pm \frac{{{\gamma ^2} + \alpha \eta + {H^2}\gamma \lambda }}{{{H^2}\gamma \lambda }},
\end{align}
for the SDW with $\mathbf{Q}_1$ and $\mathbf{Q}_2$ respectively. The critical field for the formation of SDW at $\theta_H=0$ is determined by the condition $\gamma^2+\alpha\eta+H^2\gamma\lambda=0$. For a field slightly above the critical field, we can expand $\gamma^2+\alpha\eta+H^2\gamma\lambda \approx \alpha_0[H-H_{\mathrm{SDW}}(\theta_H=0)]$ . For a weak hysteresis $G\ll 1$ observed in $\mathrm{CeCoIn_5}$, we can neglect the dependence of $H_{\mathrm{SDW}}$ on $\theta_H$. The width of the hysteretic region is
\begin{align}\label{eq9}
\Delta {\theta _H} = \frac{{{\alpha _0}\left[ {H - {H_{\mathrm{SDW}}}} \right]}}{{{H_{\mathrm{SDW}}}^2\gamma \lambda }},
\end{align} 
and it depends linearly on magnetic field. The linear dependence is guaranteed by the second-order phase transition from the nonmagnetic phase to the SDW phase. Away from the hysteretic region, there is only one SDW phase, while in the hysteretic region, two SDW states can coexists. Here, the switching of $\mathbf{Q}$ of SDW by the magnetic-field direction is of the first order, while the transition from the nonmagnetic state into the SDW phase at $H_{SDW}$ is of the second order.

\begin{figure}[t]
\psfig{figure=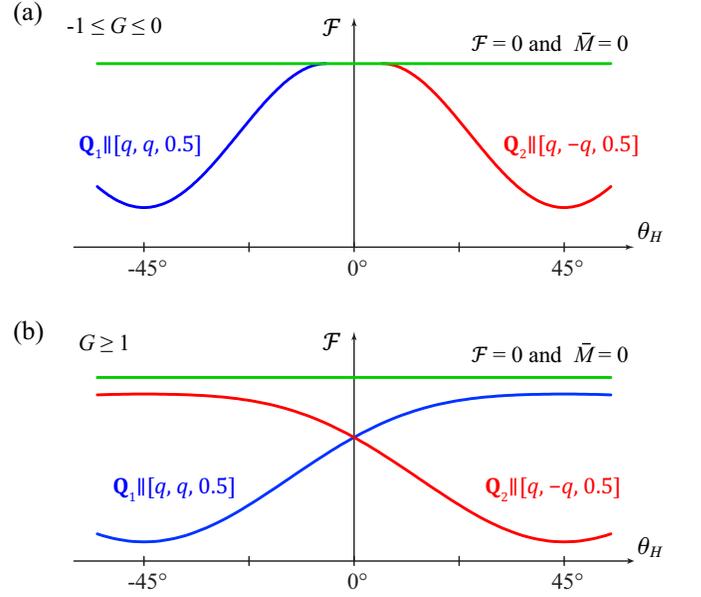,width=\columnwidth}
\caption{(color online)  Schematic view of free energy density $\mathcal{F}_{1,2}$ as a function of field angle $\theta_H$ for (a) $-1\le G \le 0 $ and (b) $G\ge 1$. The green line represents a nonmagnetic state, where the SDW gap vanishes.
} \label{f2}
\end{figure}

2. For $-1 \le G\le 0$, the switching from the SDW state with $\mathbf{Q}_1$ to the SDW state with $\mathbf{Q}_2$ is via a nonmagnetic state $M_0=0$ around $\theta_H=0$, see Fig. \ref{f2} (a). The switching involves two continuous phase transitions and there is no hysteresis. The system is always in a single domain.

3. For $G\ge 1$, there exist two minima in the free energy, corresponding to SDW states with $\mathbf{Q}_1$ and $\mathbf{Q}_2$, see Fig. \ref{f2} (b). In equilibrium, there are two coexisting SDW domains. Rotation of field direction changes the relative populations of two domains.

4. For $G\le -1$, there is no SDW phase.\\

(b) We now discuss a possibility of a double-${Q}$ solution, i.e., homogenous coexistence of SDW with $\mathbf{Q}_1$ and $\mathbf{Q}_2$ in a single domain. The solution can be written as ${M_z} = \frac{{{\bar{M}}}}{{\sqrt 2 }}\left[ {\sin \left( {{\mathbf{Q}_1}\cdot \mathbf{r}+\varphi} \right) + \sin \left( {{\mathbf{Q}_2}\cdot \mathbf{r}} \right)} \right]$. The corresponding free energy density is
\begin{align}\label{eq10}
\begin{split}
{{\cal F}_{2Q}} = - \frac{\alpha }{4}\bar{M}^2 + \frac{5 \beta }{64} \bar{M}^4 \\
-\sum_{i=1,2}\left[ \gamma \frac{{Q_i^2 }}{4}- \frac{\eta }{4}\left( {Q_{ix}^4 + Q_{iy}^4 } \right) + \frac{\lambda }{4} {{{\left( {\mathbf{H} \times {\mathbf{Q}_i}} \right)}^2} } \right]\bar{M}^2 .
\end{split}
\end{align}
The double-${Q}$ solution considered has higher energy because the coefficient of the quartic term is smaller than that of the single-${Q}$ solution in Eq. \eqref{eq2}. Moreover $\mathbf{Q}_1$ and $\mathbf{Q}_2$ are not the linearly independent optimal wave vectors due to the presence of the $\lambda$ term, which also increases the free energy. Therefore, the coexistence of two SDW states with $\mathbf{Q}_1$ and $\mathbf{Q}_2$ is not favored in the vicinity of the phase boundary. Nevertheless, the analysis does not exclude a possible double-${Q}$ solution in the nonlinear region where the SDW order parameter is large.\\

\noindent
\textbf{Thermal conductivity measurements} --
The switching of $\mathbf{Q}$, observed by neutron scattering \cite{gerber_switching_2014}, induces a discontinuous change in the thermal conductivity \cite{Kim2016}. Therefore, the details of the hysteretic nature of the domain switching can also be studied experimentally via thermal conductivity measurements. The phenomenological model above provides the theoretical background for the experiment.

\begin{figure}[t]
\psfig{figure=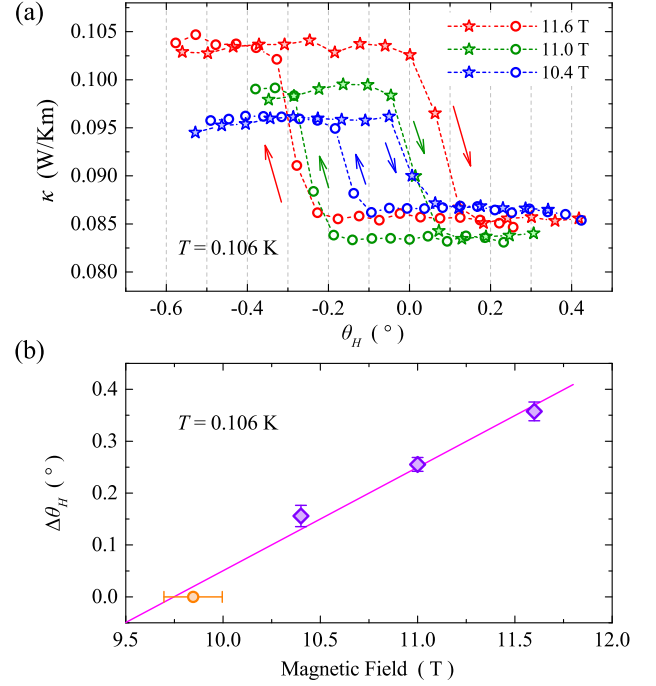,width=\columnwidth}
\caption{(color online) (a) Thermal conductivity ($\kappa$) of $\mathrm{CeCoIn_5}$ as a function of the magnetic-field direction ($\theta_H$) for three values of the field intensity at 0.106 K. (b) The width of the hysteresis region as a function of the magnetic-field intensity (purple diamonds). The widths and error bars are decided by fitting the transition regions, three data points around each step in thermal conductivity in (a), with parallel lines. The orange circle represents the $H_{SDW} = 9.8 \pm 0.15$ T at $T\approx$100 mK obtained by neutron scattering (Fig. 3 in Ref. \cite{PhysRevLett.104.127001}). The magenta line is the linear fit to the four points shown in the figure. 
} \label{f3}
\end{figure}

The thermal conductivity measurements were performed on a single crystal $\mathrm{CeCoIn_5}$ with a heat current applied along the [110] crystallographic direction, which is the nodal direction of the $d_{x^2-y^2}$-wave superconducting state in $\mathrm{CeCoIn_5}$. The thermal conductivity cell was mounted on a piezoelectric rotator with a horizontal axis of rotation. A standard one-heater and two-thermometer method was used for the measurements. The sample was oriented with the $c$ axis parallel to the rotation axis, ensuring that the vertical magnetic field, provided by the superconducting magnet, lay within the $a$-$b$ plane during the sample's rotation.

The thermal conductivity data in the vicinity of the switching transition around $\mathbf{H}\parallel$ [100] are displayed in Fig. \ref{f3}(a) for several values of the magnetic field at temperature of 0.106 K. Waiting for an equilibration after a rotation of the field and averaging to obtain high-resolution data required approximately 1 hour for each data point. The sharp jump of thermal conductivity originates from the switching of $\mathbf{Q}$ between being parallel and perpendicular to the heat current. Figure \ref{f3}(b) shows the width of the hysteresis as a function of field for the data in Fig. \ref{f3}(a), demonstrating a linear dependence on the magnetic field, consistent with the theoretical results above. Rigorously, the Landau description is valid close to the critical field $H_{\mathrm{SDW}}$. According to the neutron scattering \cite{gerber_switching_2014}, the scaling relation $M_z\sim \sqrt{H-H_{\mathrm{SDW}}}$, predicted by the Ginzburg-Landau theory, holds up to the upper critical field. This implies that the Landau description is valid for the entire $Q$ phase.

The hysteresis window decreases with increasing temperature \cite{Kim2016}. This could be caused by the suppression of the SDW gap with elevated temperature. The suppression of hysteresis is also expected in conventional first-order phase transitions due to thermal fluctuations.\\

\noindent
\textbf{Discussion} --
We start with a free energy functional based on the local magnetization density, which describes the condensation of magnetic excitations in the nodal directions of $d$-wave pairing symmetry. The magnetic field couples to the propagation vector $\mathbf{Q}$ of the SDW due to the spin-orbit interaction. Therefore, the external field lifts the degeneracy of the two equivalent directions of $\mathbf{Q}$ associated with the $d$-wave order parameter and enforces one direction for $\mathbf{Q}$. As the direction of $\mathbf{Q}$ is confined to the nodal direction of the $d$-wave order parameter, a continuous rotation of $\mathbf{Q}$ in response to the rotating magnetic field is not possible. Thus, the switching of $\mathbf{Q}$ must be a discontinuous process, and a hysteresis is expected. We argue that the switching of $\mathbf{Q}$ involves closing of the SDW gap at one $\mathbf{Q}$ and then reopening the gap at another (perpendicular) $\mathbf{Q}$. Away from the hysteretic region, the energy of the SDW with disfavored $\mathbf{Q}$ is higher than that of the nonmagnetic state. Because the closing of the SDW gap is a second-order phase transition, the disfavored SDW cannot exist, and the system has only one single SDW domain with $\mathbf{Q}$ as perpendicular as possible to the magnetic field. In the hysteretic region, there are two energy minimal states, with one being the local minimum (SDW with disfavored $\mathbf{Q}$) and the other being global minimum (SDW with favored $\mathbf{Q}$). Experimentally, however, both neutron-scattering and thermal-conductivity data do not show evidence for domains with both $\mathbf{Q}$'s. Because of thermal fluctuations/quantum tunneling, there may exist domains of SDW with two different $\mathbf{Q}$'s. The disfavored SDW domain is eliminated through suppression of the SDW gap. 

Thermal conductivity decreases with increasing $\bar{M}$ of the SDW state \cite{Kim2016}. In Fig. \ref{f3}, thermal conductivity is nearly  constant in the hysteretic region and then changes sharply during the switching. This means that $\bar{M}$ is constant in the hysteretic region as well, drops sharply to zero, and then immediately to the original $\bar{M}$ of the second $\mathbf{Q}$ during the switching process. Alternatively, the existence of multiple domains and scattering of quasiparticles by domain walls could result in nearly constant thermal conductivity before switching in the hysteretic region. To describe the sharp change of $\bar{M}$, one needs to include higher order terms in the free energy expansion. However, the qualitative picture remains valid. We stress that $\bar{M}$ decreases continuously to zero during the switching in the present picture.

To summarize, we have studied the dynamics of switching of the ordering wave vector $\mathbf{Q}$ of the SDW state in $\mathrm{CeCoIn_5}$. We argue that, in the course of switching, the gap of the SDW at one $\mathbf{Q}$ is closed, and immediately the gap of the SDW at a perpendicular $\mathbf{Q}$ opens. We provide a simple phenomenological model to describe this hysteretic process. The hysteresis window is shown to grow linearly with the magnetic field, which is consistent with the experiments. The agreement between theory and experiments supports our scenario of the hypersensitivity of the $Q$ phase on the direction of magnetic field, as well as the magnon condensation as the origin of the SDW phase.

\begin{acknowledgments}
The work was carried out under the auspices of the U.S. DOE Contract No. DE-AC52-06NA25396 through the LDRD program. The experimental work was supported by the U.S. Department of Energy, Office of Basic Energy Sciences, Division of Materials Sciences and Engineering.
\end{acknowledgments}

\bibliography{reference}

\end{document}